\documentclass[epjST]{svjour}
\usepackage{graphics,soul}
\usepackage{graphicx,amssymb,amsmath,subfigure}
\usepackage[latin1]{inputenc}
\usepackage{color}

\definecolor{red}{rgb}{1,0,0}
\definecolor{green}{rgb}{0,1,0}
\definecolor{blue}{rgb}{0,0,1}
\definecolor{darkmagenta}{rgb}{.5,0,.5}

\begin{document}

\title{The role of intermediaries in the synchronization of pulse-coupled oscillators. }

\author{Rodrigo A. Garc\'ia\inst{1} \and Nicol\'as Rubido\inst{2,1} \and
  Arturo C. Mart\'i\inst{1} and Cecilia
  Cabeza\inst{1}\fnmsep\thanks{\email{cecilia@fisica.edu.uy}} }

%
\institute{Facultad de Ciencias, Universidad de la Rep\'ublica, Igu\'a
  4225, Montevideo, Uruguay \and Institute for Complex Systems and
  Mathematical Biology, University of Aberdeen, King's College, AB24
  3UE Aberdeen, UK}
\abstract{The role of intermediaries in the synchronization of small groups of light
controlled oscillators (LCO) is addressed. A single LCO is a two-time-scale phase oscillator.
When pulse-coupling two LCOs, the synchronization time decreases monotonously as the coupling
strength increases, independent of the initial conditions and frequency detuning. In this
work we study numerically the effects that a third LCO induces to the collective behavior
of the system. We analyze the new system by dealing with directed heterogeneous couplings
among the units. We report a novel and robust phenomenon, absent when coupling two LCOs,
which consists of a discontinuous relationship between the synchronization time and coupling
strength or initial conditions. The mechanism responsible for the appearance of such
discontinuities is discussed.
} 
\maketitle
\section{Introduction}
\label{intro}

Collective behavior, such as synchronization \cite{pikovsky2003synchronization}, is
ubiquitous in nature. In particular, a paradigmatic example of synchronous behavior is
observed in gregarious fireflies communities  \cite{buck1988synchronous}. Initially inspired
in the emissions of gregarious fireflies,  light controlled oscillators (LCOs) are  aimed to
study the mechanisms which drive their collective behavior into synchronization
\cite{ramirez2003synchronization,rubido2009experimental,rubido2011synchronization,rubido2011scaling,avila2011firefly}.

An LCO is a relaxation oscillator which interacts with others by means of light pulses. When
an LCO is uncoupled, its oscillation is composed of a charging state and a discharging state.
Usually, the characteristic time-scales are chosen such that the charging state lasts
significantly more than the discharging state (an extreme case is the integrate-and-fire
oscillation, where discharge is instantaneous). When considering two or more coupled LCOs,
the dynamics of the individual oscillators is modified due to the interaction. The reason
is that, during the discharging state, the LCO is able to emit a pulse which affects the
LCOs connected to it. The influence of the received pulse over the oscillation is either
excitatory (the oscillation is accelerated), if the affected LCO is at its charging state,
or inhibitory (the oscillation is decelerated), if the affected LCO is at its discharging
state.

One of the advantages of the LCOs is that they are easily implemented using simple electronic
devices \cite{ramirez2003synchronization,rubido2009experimental} that facilitate the
experimental analysis of their transient times and emergent properties. The transient
time needed to synchronize, i.e., the synchronization time, is of crucial interest in several
technological fields. In general, when two self-sustained oscillators are coupled, the
synchronization time deceases as the coupling strength is increased. In particular, this
behavior is also observed between two interacting LCOs. However, in the special case of
bidirectional coupling between two LCOs, a critical coupling strength value exists beyond
which any higher value of coupling destroys the oscillation. Such a novel phenomenon is known
as oscillation death \cite{ghosh2012design,banerjee2012enhancing,hens2013oscillation}.

In the case when the interaction between two LCOs is asymmetrical, the oscillators can be
ordered according to a well-defined hierarchy. Namely, the oscillator that influences the
most is said to have the highest hierarchy. For example, this is the case of a
\textit{master-slave} (MS) configuration, where one LCO influences another without being
influenced by it. In contrast, when the coupling is bidirectional, with no higher or lower
hierarchy, the system is said to be in a \textit{mutual interaction} (MI) configuration.
Ref.~\cite{cosenza2010equivalent} exploides a parallelism between  MS and MI configurations and  driven and autonomous  maps
which display equivalent synchronization characteristics.

When dealing with the task of lowering the synchronization time between two oscillators in
MS configuration, a possibility is to add a third oscillator that mediates between the master
and the slave, i.e., an intermediate hierarchy. On the other hand, if an intermediate
hierarchy is added to the MI configuration, the directionality of the added links destroys
the equality in hierarchies of the original LCOs configuration and the resultant
configuration is a non-trivial hierarchical relationship. 

The study of synchronization in systems of three oscillators has been addressed in previous works, for example, relay coupling \cite{gutierrez2013generalized},
environmental coupling \cite{resmi2010synchronized}, or chemical oscillators \cite{wickramasinghe2011phase}. In those cases, the oscillators are arranged in a three-on-a-row 
configuration and, as a result, the interaction between the \textit{master} and the \textit{slave} is given exclusively through the intermediary. 
While in those cases the equations of motion are continuous and have continuous derivatives, we deal here with piecewise analytical equations of motion, 
which introduce singularities in the derivatives.

In this work we study the influence of adding an intermediary oscillator to the
synchronization times of two pulse-coupled oscillators using various coupling
configurations. Our numerical experiments show the appearance of a novel phenomenon
in the relationship between the coupling strength and synchronization time. Contrary to
the monotonous decrease that the synchronization times of two oscillators exhibit as the
coupling strength increases, the novel phenomenon is accounted by a discontinuous
relationship between synchronization times and coupling strengths and is solely explained
due to the inclusion of the intermediary. Moreover, we derive analytically the conditions
the intermediary oscillator must fulfil in order to reduce the transient times towards
synchronization.

 \section{Model}
The equation describing the evolution of the $i$-th coupled LCO
is \cite{ramirez2003synchronization,rubido2011synchronization}
\begin{equation}
 \frac{dV_{i}(t)}{dt}=\left[E_{i}\left(t\right)-\lambda_{i}\left(V_{i}\left(t\right)-V_{cc}
  \right)\right]\epsilon_{i}+\left[E_{i}\left(t\right)-\gamma_{i}V_{i}\left(t\right)\right]
   \left(1-\epsilon_{i}\right),
 \label{eq:Sistema din=0000E1mico}
\end{equation}
where $V_i(t)$ is the oscillating state variable of the $i$-th oscillator at time $t$,
$\epsilon_{i}$ is a binary value that represents the state in which the $i$-th oscillator
is, taking the value $1$ for the \emph{charging state} and $0$ for the \emph{discharging
state}, the parameter $V_{cc}$ determines the oscillation amplitude, and $E_i(t)$ is the
coupling. The state of every LCO changes from charge (discharge) to discharge (charge),
$\epsilon_i = 1\to0$ ($\epsilon_i = 0\to1$), when its state variable, $V_i(t)$, reaches
$2V_{cc}/3$ ($V_{cc}/3$). The parameters $\lambda_{i}$ and $\gamma_{i}$ define the
characteristic charging and discharging state frequencies, respectively. Hence, the $i$-th
LCO natural (uncoupled) frequency is
\begin{equation}
  \nu_i = \left(T_{\lambda_i} + T_{\gamma_i}\right)^{-1} = \left(\frac{\log 2}{\lambda_i}
   + \frac{\log 2}{\gamma_i}\right)^{-1}\,.
 \label{eq_LCO_freq}
\end{equation}

The choice of notation in Eq.~(\ref{eq:Sistema din=0000E1mico}) and the dimensional values
for the LCOs (units of voltage for $V_i(t)$, frequency for $\lambda_i$ and $\gamma_i$, and
units of voltage per second for the $E_i(t)$) are inspired in the experimental implementation
of the system \cite{rubido2009experimental}. This experimental setup is based on a dual RC
circuit, with a charging state given by the characteristics of one of the RC and a
discharging state given by the characteristics of the second RC circuit.

The coupling term is explicitly given by
\begin{equation}
  E_{i}\left(t\right)=\underset{j\neq i}{\sum}a_{ij}\beta_{ij}\left(1-\epsilon_{j}\right),
 \label{eq:Acoplamiento}
\end{equation}
$a_{ij}$ being the adjacency matrix entry corresponding to the possible connection between
nodes $i$ (LCO$_i$) and $j$ (LCO$_j$), namely, $a_{ij} = 1$ if the LCOs are connected and
$a_{ij} = 0$ otherwise, and $\beta_{ij}$ being the  coupling strength of the particular
connection. The state modification that is induced by this coupling function results in an
acceleration of the charging state (excitation) and/or a deceleration for the discharging
state (inhibition). Asymmetry in the couplings is achieved by setting $\beta_{ij}\neq
\beta_{ji}$, though the adjacency matrix is assumed to be always symmetrical. For example,
an MS configuration between LCOs $i$ and $j$ is achieved by setting $\beta_{ij} > 0$ and
$\beta_{ji} = 0$.

\begin{figure}[ht]
 \begin{center}
  \includegraphics[width=.45\columnwidth]{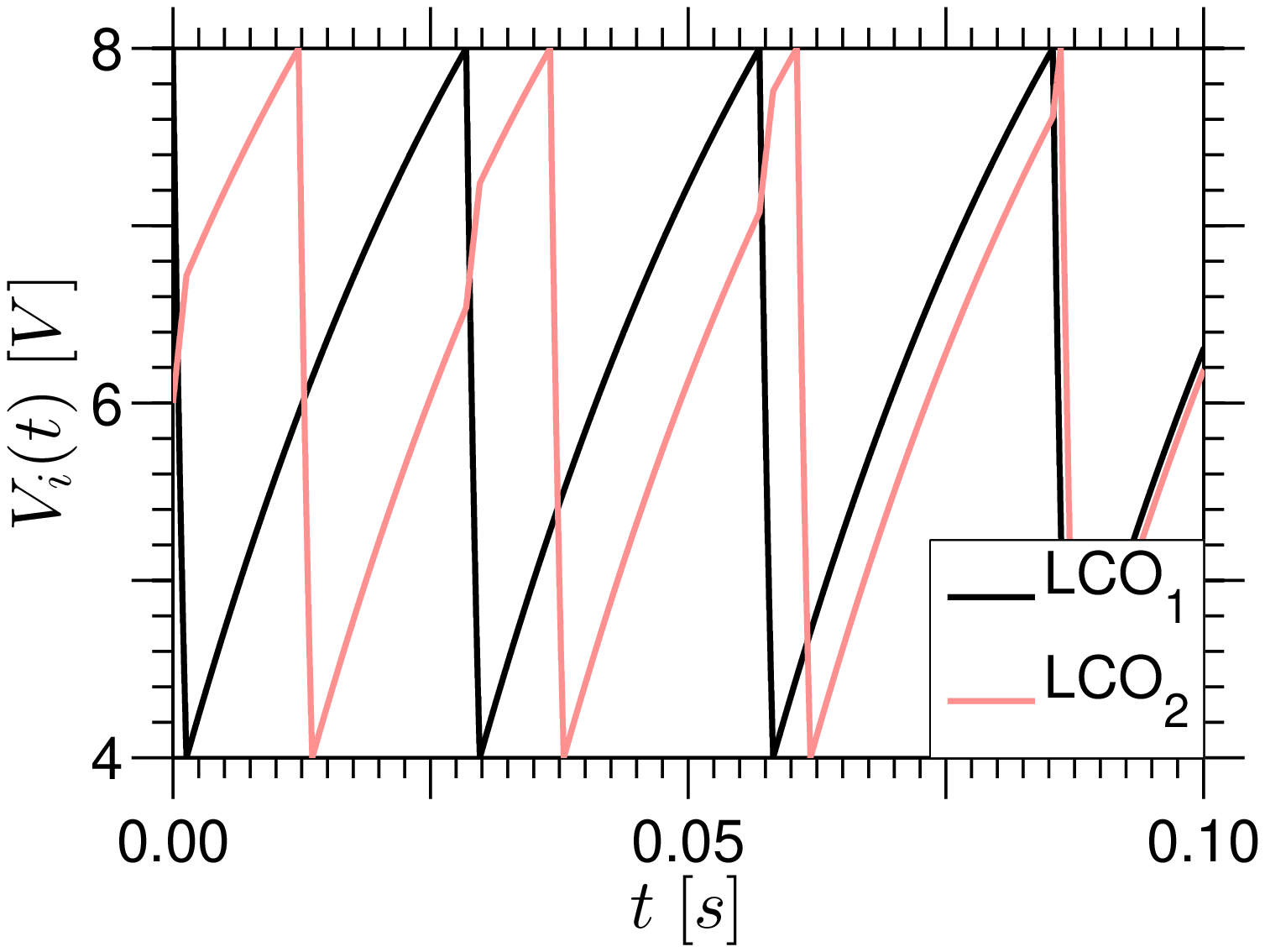}
  \includegraphics[width=.45\columnwidth]{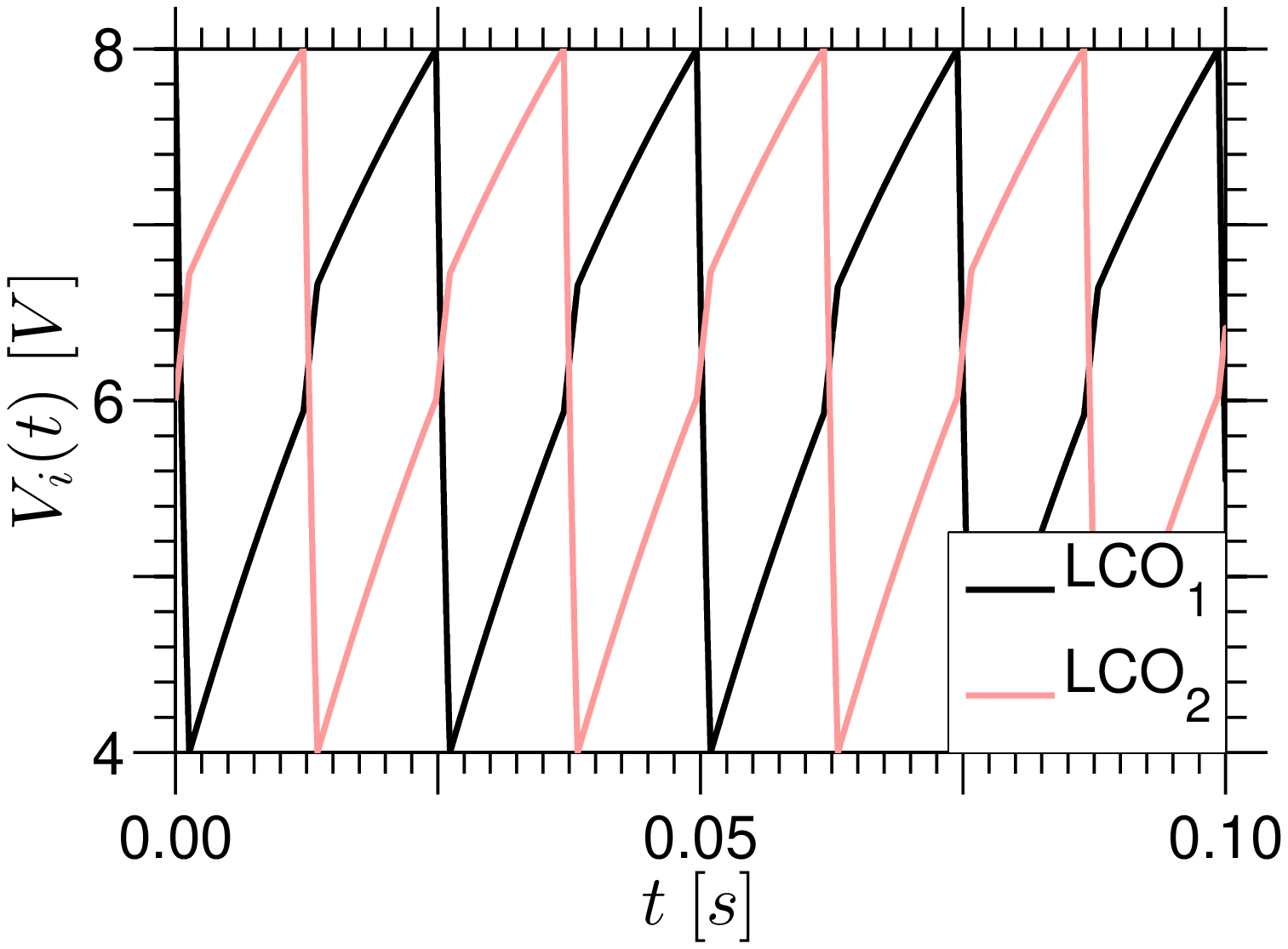}
 \end{center}
\caption{Evolution window of the state variable $V_i(t)$ of two LCOs
coupled in a MS configuration (left) and MI configuration (right). In both
panels the oscillators are identical, with $T_{\lambda}=27.8\;ms$ and
$T_{\gamma}=0.7\;ms$, and the coupling strength $\beta = 650\;V/s$.}
\label{fig:timeseries}
\end{figure}

Although Eq.~(\ref{eq:Sistema din=0000E1mico}) is a piecewise linear coupled set of
differential equations, a general and unique analytical solution, which is piecewise smooth,
exists for each initial condition \cite{rubido2011synchronization}. In particular, a solution
for two coupled LCOs in a MS (left panel) and MI (right panel) configuration are shown in
Fig.~\ref{fig:timeseries} using the same initial conditions ($V_1(0) = 2V_{cc}/3$ discharging
and $V_2(0) = V_{cc}/2$ charging, with $V_{cc} = 12$~V). We note that the initial condition
choice represents the unstable anti-synchronous state of the MI configuration (i.e., both
oscillators perform the same oscillation but with a phase difference of $\pi$), hence, the
behavior of the transient time-window shown in the right panel of Fig.~\ref{fig:timeseries}.

As each LCO is a one dimensional oscillator, piecewise linear, a phase $\phi_{i}\left(
V_{i}\right)$ is defined using the free oscillation frequency of each LCO. Specifically,
\begin{equation}
 \phi_{i}\left(V_{i}\right) = \begin{cases}
2\pi \left(   \frac{ -1 }{ \log 2 }\frac{ \gamma_i }{ \lambda_i + \gamma_i }
   \log\left[\frac{3}{2}\left( 1 - \frac{ V_{i}\left(t\right) }{ V_{cc} } \right)\right]  \right) \,,
    & \epsilon = 1\,, \\
 2\pi\left( 1 - \frac{ 1 }{ \log 2 }\frac{ \lambda_i }{ \lambda_i + \gamma_i }
  \log\left[ 3\frac{ V_{i}\left(t\right) }{ V_{cc} }\right]\right)\,,
    & \epsilon = 0\,. \end{cases}
 \label{eq_LCO_phases}
\end{equation}

\begin{figure}[htbp]
 \begin{center}
  \includegraphics[width=.6\columnwidth]{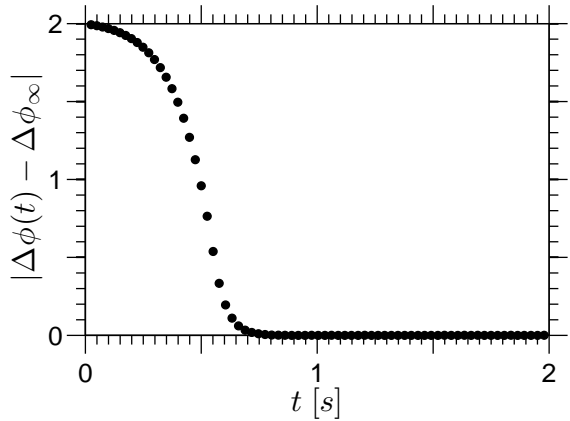}
 \end{center}
\caption{Phase difference evolution towards synchronization between a \textit{master}
LCO and a \textit{slave} LCO in the interaction between two oscillators. The absolute value of the difference between the phase
difference and the final phase difference decreases to zero, and it remains constant when
the synchronous state is reached. In this case the synchronization time is $804.8\;ms$.}
\label{fig_phase_dif}
\end{figure}

Equation~(\ref{eq_LCO_phases}) allows to determine the phase difference evolution, namely,
$\Delta \phi(t)$, between LCOs. This provides a quantitative measure of the system's emergent
properties. For the synchronization, it is expected for it to converge to a constant value
(zero if the system has complete synchronization \cite{lu2005adaptive} and finite for phase
synchronization \cite{rosenblum1996phase}). Hence, in order to calculate the synchronization
times, the difference, $\Delta(t)$, between $\Delta \phi(t)$ and its limit value, $\Delta
\phi_\infty = \lim_{t\to\infty} \Delta \phi(t)$, is defined. For example,
Fig.~\ref{fig_phase_dif} shows how $\Delta(t)$ decreases as the system evolves towards the
synchronized state, regardless if it is a complete synchronous ($\Delta \phi_\infty = 0$)
state or a phase synchronous ($\Delta\phi_\infty > 0$) state. In both cases, $\lim_{t\to\infty}
\Delta(t) = 0$.

\section{Results }
\label{sec:results}

\subsection{Two-LCO dynamics.}
\label{ss:results2}
We start the study of synchronization times restricting ourselves to the case of two
coupled LCOs. The possible hierarchies in such cases are the \emph{master-slave} (MS)
configuration (extreme hierarchical difference) and the \emph{mutual interaction} (MI)
configuration with symmetrical coupling strengths (identical hierarchy). For the sake of
simplicity, we deal with identical oscillators, namely, $\lambda_i = \lambda$, $\gamma_i
= \gamma$, and $\beta_{ij} = \beta$, for all LCOs from $i = 1,\,2$. In order to construct
the synchronization time dependence on the coupling strength, the initial condition of one
of the oscillators is set to be $V_1(0) = 2V_{cc}/3$ (which corresponds to the master LCO
in the MS configuration), and  the other oscillator is set to be $V_2(0) = V_{cc}/2$.

We note that, unless the initial conditions for the oscillators are set to be identical
[$V_1(0) = V_2(0)$], the synchronization time, $t_{s}$, decreases monotonously as the
coupling strength ($\beta$) is increased, regardless of the particular choice of initial
condition. The reason for such universal behavior to exist is that the system,
Eq.~(\ref{eq:Sistema din=0000E1mico}), is scalable into a non-dimensional form which
only depends on the ratio between $\lambda$ and $\gamma$ and the coupling strength $\beta$
\cite{rubido2011scaling}. The atypical case, $V_1(0) = V_2(0)$, corresponds
to the synchronization manifold \cite{pecora1998master} of the coupled system when the LCOs have identical parameters
($\lambda$ and $\gamma$), hence, its synchronization time vanishes for any coupling strength.
In other words, for any initial condition that avoids the case $V_1(0) = V_2(0)$, we
observe an scalable curve for the synchronization times as a function of coupling strength.

\begin{figure}[htbp]
 \begin{center}
  \includegraphics[width=.75\columnwidth]{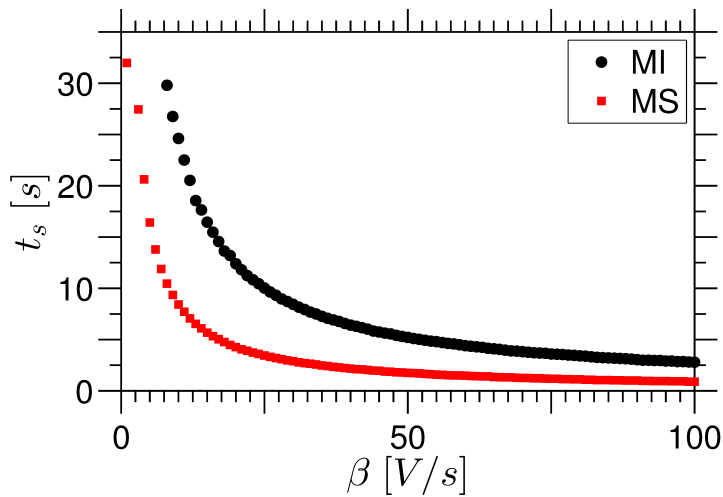}
 \end{center}
\caption{Synchronization time as a function of the coupling strength for {\it master-slave}
(MS, filled light --red online-- squares) and {\it mutual interaction} (MI,
filled dark --black online-- circles) configurations. In both cases, the two
oscillators are identical, with a period given by $T_{\lambda}=27.8$ ms and
$T_{\gamma}=0.7$ ms [Eq.~(\ref{eq_LCO_freq})]. Initial conditions are fixed to
$V_{1}(0)=2V_{cc}/3$ (the \textit{master} in the MS configuration) and
$V_{2}(0)=V_{cc}/2$ (the \textit{slave} in the MS
configuration). } \label{fig_3}
\end{figure}

An example of the universal behavior synchronization times exhibit for both 
configurations, MI (filled dark --black online-- circles) and MS (filled light --red
online-- squares), is shown in Fig.~\ref{fig_3}. This figure shows that the synchronization
time decreases monotonously as the coupling strength increases. Moreover, we observe that,
for every given $\beta$, a lower synchronization time is achieved in the MS configuration in
comparison to the MI configuration for this particular choice of initial conditions.
Furthermore, at sufficiently weak coupling strengths ($\beta\lesssim10\;V/s$), a
non-synchronous region appears for the MI configuration which is absent in the MS case
(left region in Fig.~\ref{fig_3}). The reason for the existence of such region in the MI
case is that  the choice of initial conditions place the system in the
anti-synchronous manifold, hence, its synchronization time its strictly infinite, though
any perturbation draws the system out of this manifold allowing the system to achieve
synchrony only at very large times.

When coupling two non-identical LCOs, a complete synchronization manifold is absent
\cite{acharyya2012synchronization}. Nevertheless, we observe a similar universal behavior
for the monotonous decrease in the synchronization times as the coupling strength increases.
The reason is that the system is able to synchronize with a phase lag, i.e., phase
synchronization \cite{pikovsky2003synchronization}, as long as the parameter difference
between the LCOs is small enough (large parameter differences drive the system  away from
the (1:1) Arnold tongue \cite{rubido2011synchronization}).

\subsection{Three-LCO dynamics.}
\label{ss:results3}
\begin{figure}[htbp]
 \begin{center}
 \subfigure[ ]{

  \includegraphics[width=.35\columnwidth]{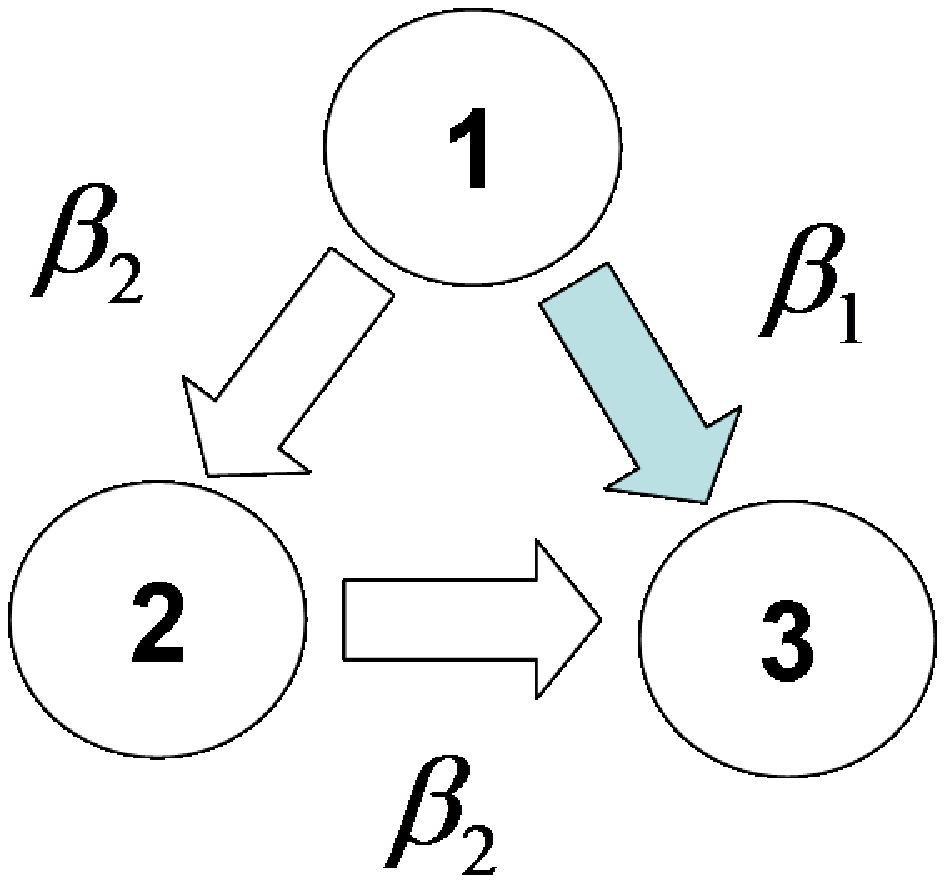} \hspace{2pc}
}
\subfigure[ ]{
  \includegraphics[width=.35\columnwidth]{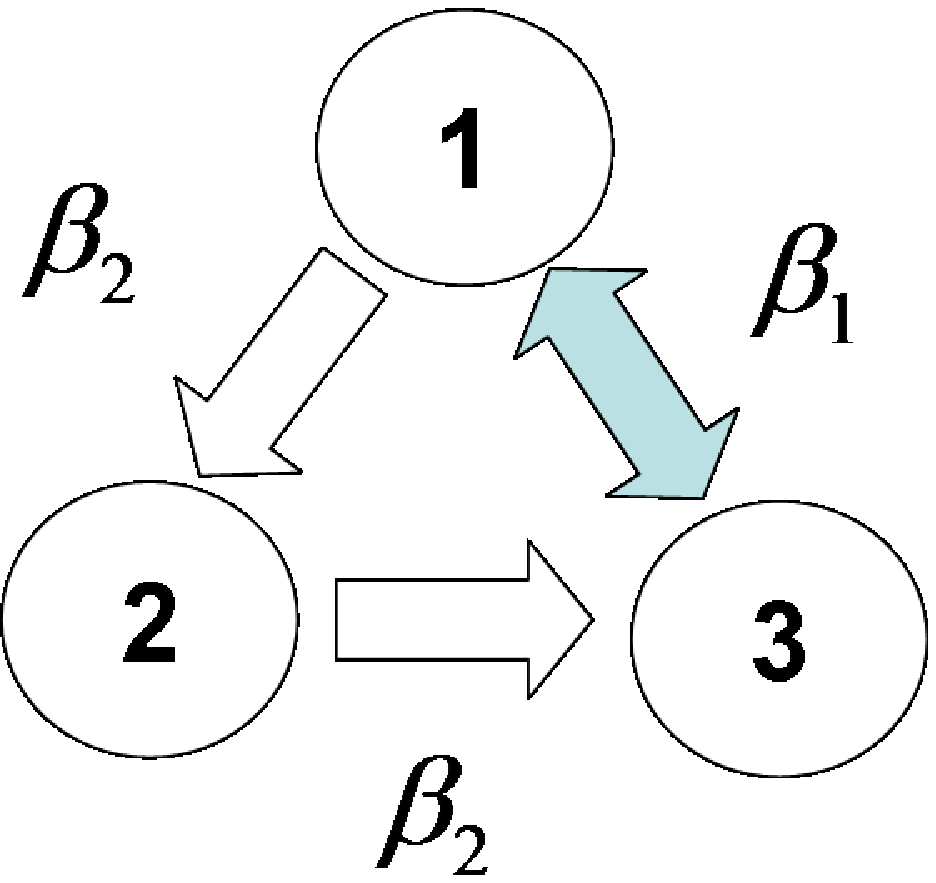}
}
  \end{center}
\caption{Schematic diagrams of the master-slave (MS) configuration [oscillators
$1$ and $2$] with an intermediary [oscillator $2$] (a) and the mutual
interaction (MI) configuration with an intermediary (b). The colored links correspond to the MS 
and MI without an intermediary oscillator. The coupling strengths involving 
the intermediary oscillator (LCO$_2$) are set to be symmetrical to
reduce the parameter space dimension.}
\label{Configurations}
\end{figure}

The role of including an intermediary oscillator on the MS (MI) configuration is
dealt by analyzing the modification on the synchronization times between the oscillators
composing the original MS (MI) configuration. In the following, we note these oscillators
as LCO$_1$ and LCO$_3$, and the added intermediary as LCO$_2$. In particular,
Fig.~\ref{Configurations} shows a scheme of how LCOs $1$ and $3$ interact (light filled
arrows) in MS (left panel) and MI (right panel) configuration, with LCO$_2$ acting as an
intermediary (unfilled arrows). Also, the schematic diagrams show the corresponding
notation for the coupling strengths  involved in each particular case.

\begin{figure}[htbp]
\includegraphics[width=.49\columnwidth]{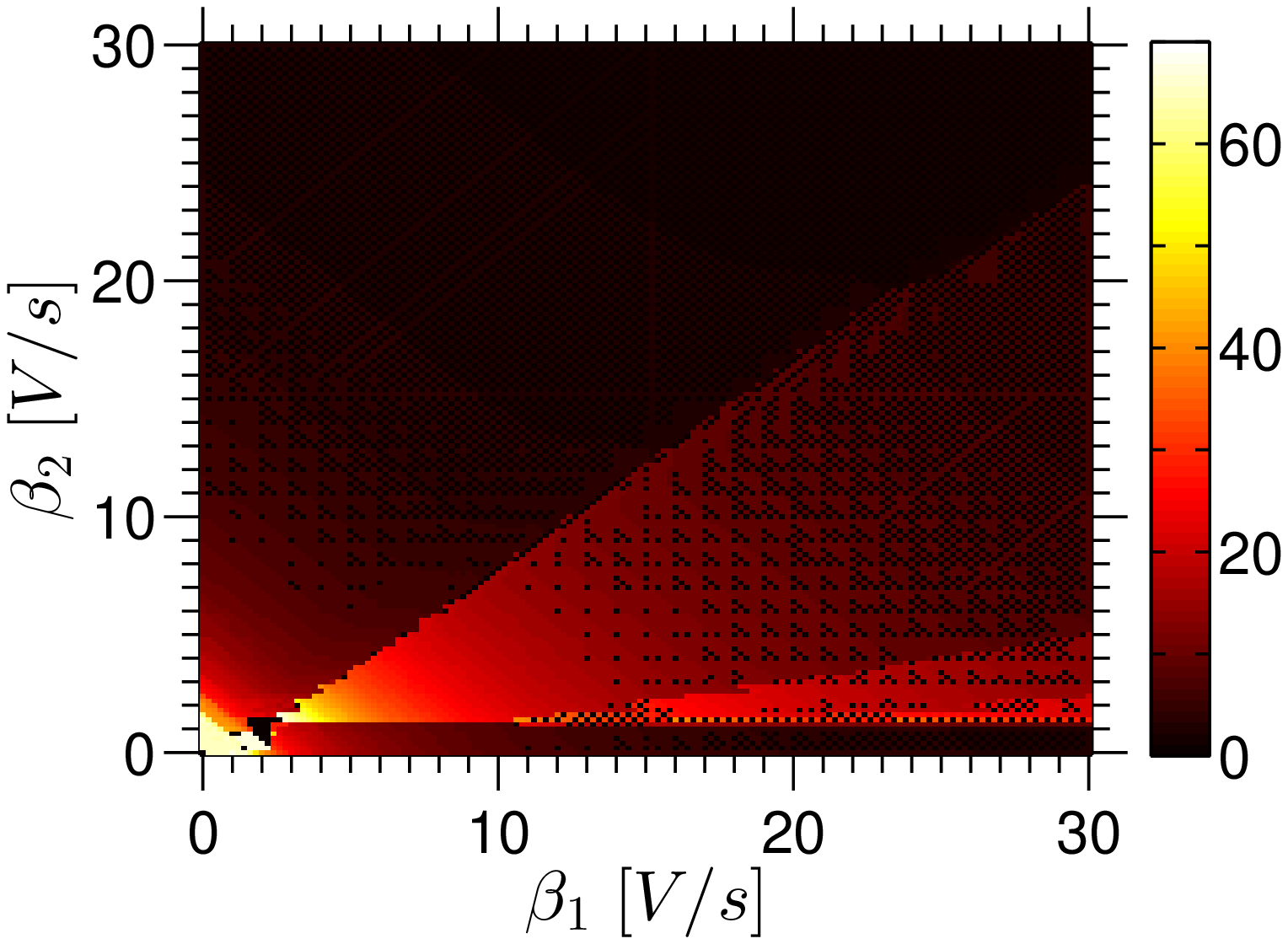}
\includegraphics[width=.49\columnwidth]{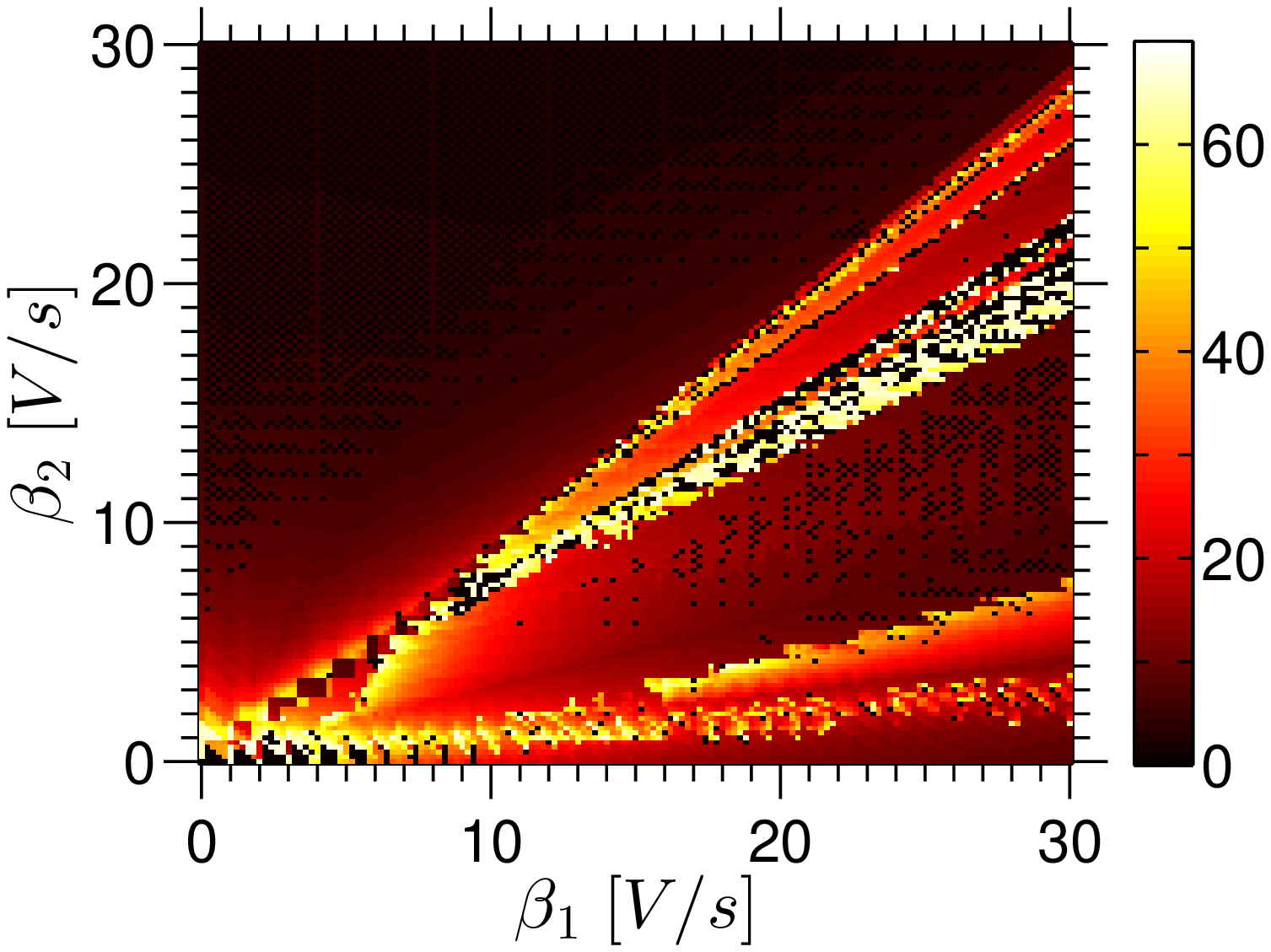}
\caption{Synchronization time (colour code) as a function of the
coupling strengths $\beta_1$ and $\beta_2$ (see Fig.~\ref{Configurations} for details
on the coupling strengths notation and configuration schematics) for the master-slave with
intermediary (left panel), and mutual interaction with intermediary (right panel)
configurations.}
\label{b1b2maps}
\end{figure}

We study the effect of the intermediary by constructing a two dimensional map of the
synchronization times as a function of $\beta_{1}$ and $\beta_{2}$, with both coupling
strengths varying from $0$ to $30\;V/s$ with a step of $0.2\;V/s$. Initial conditions
are fixed to $\vec{V_{0}}=\frac{V_{cc}}{3}\left(2,\,1.75,\,1.5\right)$ and the remaining
parameters are chosen such that the LCOs are identical ($\lambda = 26\;s^{-1}$ and
$\gamma = 1050\;s^{-1}$). Figure~\ref{b1b2maps} shows in colour code such synchronization
times in a bidimensional map for the two configurations shown in Fig.~\ref{Configurations}.

\begin{figure}[htbp]
 \begin{center}
  \includegraphics[width=.48\columnwidth]{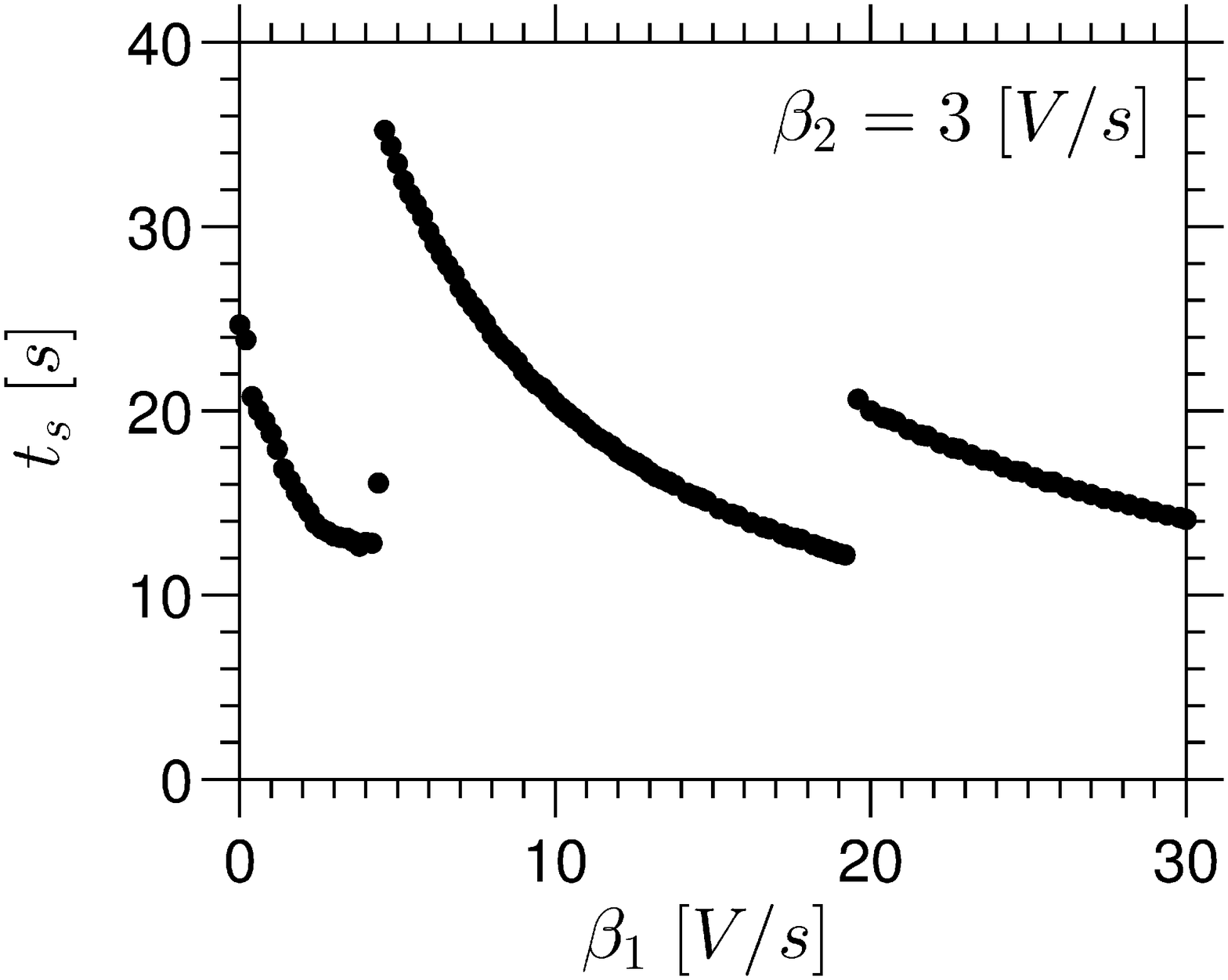}
  \includegraphics[width=.49\columnwidth]{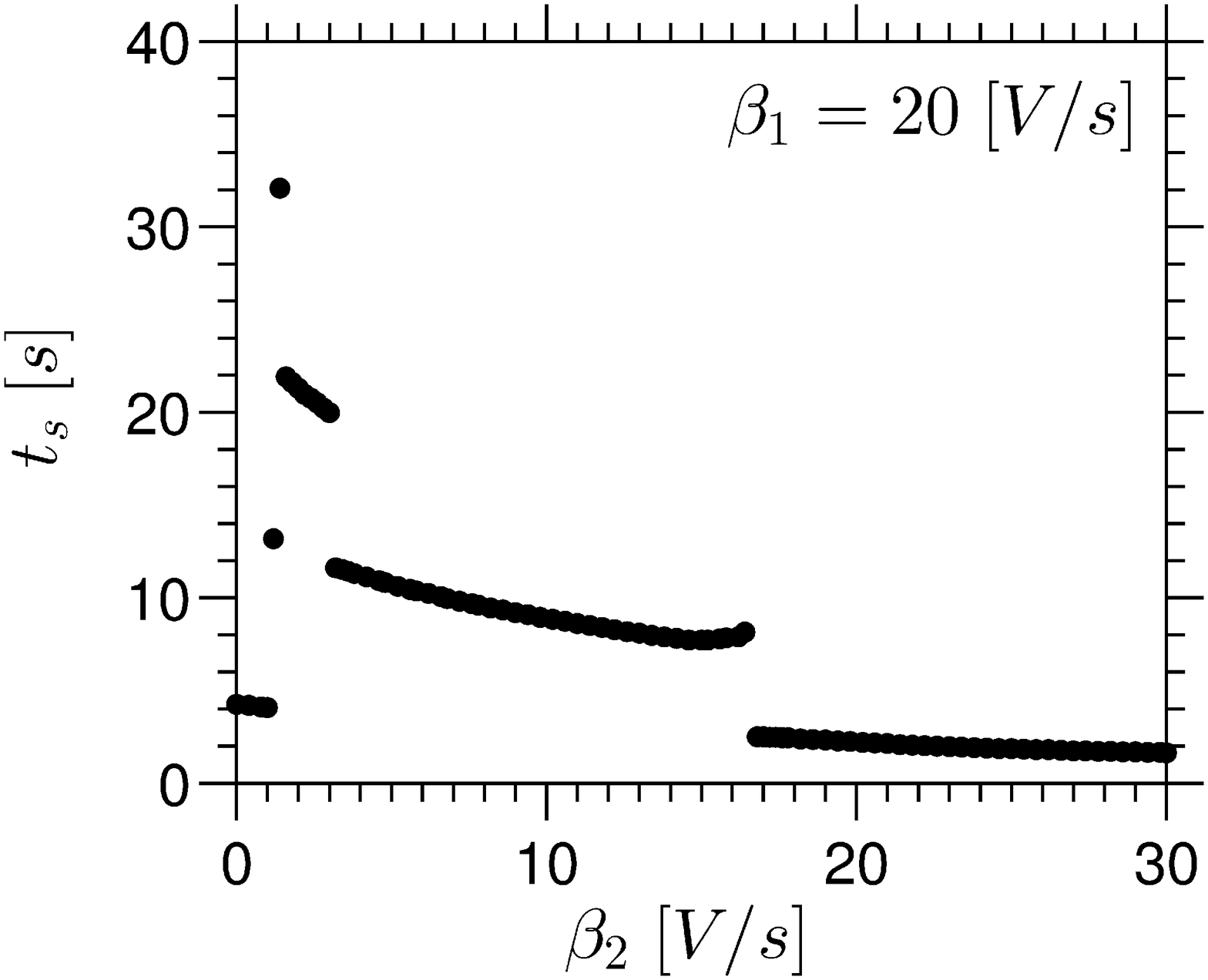}
 \end{center}
\caption{ One-dimensional horizontal (left panel) and vertical (right panel) sections
of the map depicted on the left panel of Fig.~\ref{b1b2maps} (i.e., MS with intermediary).
Discontinuities in the synchronization times are readily visualized. The section on the
left (right) panel is constructed by taking a coupling parameter value of $\beta_{2} = 3\;V/s$
($\beta_{1} = 20\;V/s$), with $\beta_{1}$ ($\beta_{2}$) varying from $0$ to $30\;V/s$ with
a step of $0.2\;V/s$. }
\label{b1b2sections}
\end{figure}

In this figure, for constant values of the $\beta_{1}$, we observe an overall trend in which
the synchronization time decreases as $\beta_{2}$ increases.  We also observe that the
relationship between synchronization times and $\beta_{1}$  is more complex for fixed
positive values of $\beta_{2}>0$ than for $\beta_{2}=0$. However, for a wide range of
values of $\beta_{1}$, the relationship between the synchronization time and the coupling
strength cease to be monotonous and, in fact, the presence of discontinuities is notorious.
Surprisingly, for some values of $\beta_1$, the synchronization time increases when
increasing the coupling strength $\beta_{2}$. The existence of discontinuities in the
relationship between the synchronization times and coupling strengths is also notorious.
This phenomenon is best illustrated by taking one dimensional sections of the map, as it is
shown in Fig.~\ref{b1b2sections}.

\section{Interpretation of the discontinuities in the synchronization times}

To gain a deeper insight in the emergent collective behavior found in the dynamics of 3
identical LCOs, we look at the effect of the master over the slave in the two-LCO dynamics.
In general, for an arbitrary phase difference, the effect of the interaction is to speed up
the evolution of the slave phase until the phase difference vanishes in the synchronous
state. The only exception is when the phase of the slave is slightly advanced in relation
with the master in such a way that both oscillators are, for some time interval,
simultaneously in their discharging state. In this case, the master delays the slave phase
until the synchronous state is reached. The limiting condition between these two cases is
given by a critical phase difference $\Delta\phi_{c}=2\pi\frac{\lambda}{\lambda+\gamma}$.
In our case, as we are dealing with relaxation oscillators (with a fast discharge), this
critical phase difference is considerably small $\Delta\phi_{c} \sim 2\pi/40$. Therefore, if
the initial phase difference is less than $\Delta\phi_{c}$ and the slave phase advanced with
respect to the master, the synchronization time is relatively small. On the opposite case,
if the initial phase difference is greater than the critical value, the master speeds up the
phase of the slave until they are synchronized with a phase difference of $ 2\pi$. In the
latter case, the synchronization time is relatively large. To summarize, in the case of two
identical LCOs interacting in a MS configuration, the relationship between the initial phase
difference and the synchronization time is discontinuous, and the synchronous state is
reached once the phase difference between both oscillators vanishes. In particular, if the
phase difference is zero for any time, it will remain zero in the future.

Next, let us consider the dynamics of three LCOs interacting in the MS configuration with
an intermediary [Fig.~\ref{Configurations}(a)], namely, configuration A. In this case, the
evolution is far more complex than in the case with only two LCOs. Here, even if the phase
difference between the master LCO$_1$ and the slave LCO$_3$ is zero for some particular time,
the implication of a stable synchronized state between these oscillators is lost. The reason
is that the intermediary, LCO$_2$ is influencing the slave, leading to a phase difference
between master and slave.

The synchronization between the master and the slave is only possible after the master is
synchronized with the intermediary. In configuration A, the intermediary is unaffected, thus,
the master-intermediary interaction is identical to a master-slave configuration in a two LCO
system and the synchronization times display similar curves to that of Fig.~\ref{fig_3}. As
mentioned before, when the synchronous state between master and slave is reached, their phase
difference will be zero and their discharging times will coincide. Henceforth, when they are
synchronized, the influence of LCO$_1$ and LCO$_2$ over the slave, LCO$_3$, can be considered
as the interaction of only one identical LCO, LCO$_{1,2}$, with LCO$_3$ in a MS configuration
but with a different coupling strength $\beta_{1}+\beta_{2}$. The initial conditions for this
equivalent interaction are the phases at the time in which the master and the intermediary
reach their synchronous state. Denoting, the synchronization time between master and
intermediary as $t_{s\,1-2}$, the initial conditions is $\text{\ensuremath{\phi}}_{1,2}^{i}=
\phi_{2}\left(t_{s\,1-2}\right)=\phi_{3}\left(t_{s\,1-2}\right)$ and $\phi_{3}^{i}=\phi_{3}
\left(t_{s\,1-2}\right)$.

In the light of the preceding analysis, the synchronization time between the master and the
slave is expressed as the sum of the synchronization time between the master and the
intermediary and the synchronization time between the equivalent LCO$_{1,2}$ and the slave
\begin{equation}
 t_{s\,1-3} = t_{s\,1-2} + t_{s\,1,2-3}
\label{eq:tiempos}
\end{equation}
where $t_{s\,1-3}$ is the synchronization time between master and slave and $t_{s\,1,2-3}$
is the synchronization time between LCO$_{1,2}$ and the LCO$_3$ in a MS configuration with
coupling strength $\beta_{1}+\beta_{2}$ and initial conditions  $\phi_{1,2}^{i}$ and
$\phi_{3}^{i}$. As the dependence of $t_{s\,1-2}$ on the coupling strength is similar to that
presented in Fig.~\ref{fig_3}, the origin of the discontinuities observed in
Figs.~\ref{b1b2maps} and \ref{b1b2sections}, lies in the second term of the r.h.s. of
Eq.~\ref{eq:tiempos}.  As the quantity $t_{s\,1,2-3}$ depends directly on the initial
conditions,  $\phi_{1,2}^{i}$ and  $\phi_{3}^{i}$, it presents a discontinuity provided that
the initial phase difference is equal to the critical phase difference $\Delta \phi_c$
previously defined. The specific values of the initial phases, $\phi_{1,2}^{i}$ and
$\phi_{3}^{i}$, depend on the coupling strength $\beta_1$ and $\beta_2$ (as they influence
$t_{s\,1-2}$), originating the mentioned discontinuities.

\begin{figure}[htbp]
 \begin{center}
  \includegraphics[width=.6\columnwidth]{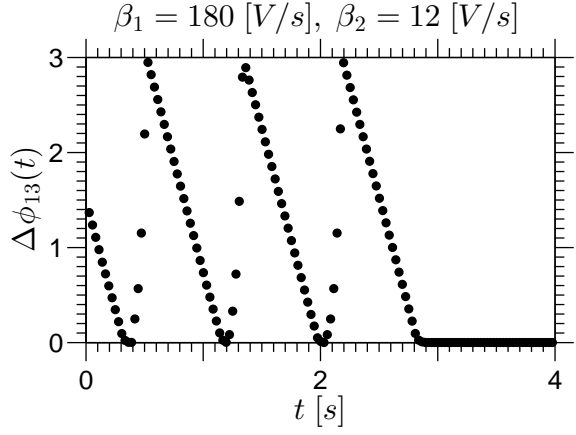}
 \end{center}
\caption{Phase difference between LCO$_1$ and LCO$_3$ for configuration A   shown in Fig. \ref{Configurations}, under the same conditions given
in Fig. \ref{b1b2maps}.  Phase of the LCO$_ 3$ is first attracted to   the phase of LCO 1, but the perturbations made by LCO$_2$ unable the
synchronization between them, attracting the phase of LCO$_3$ to the   phase of the LCO$_2$, condition that is then perturbed again by LCO$_1$. 
LCO$_1$ and LCO$_3$ are not ready to synchronize until LCO$_1$ and LCO$_2$ synchronize first, 
only then the phase difference will evolve to a   constant value. }
\end{figure}

\begin{figure}[htbp]
 \begin{center}
  \includegraphics[width=.7\columnwidth]{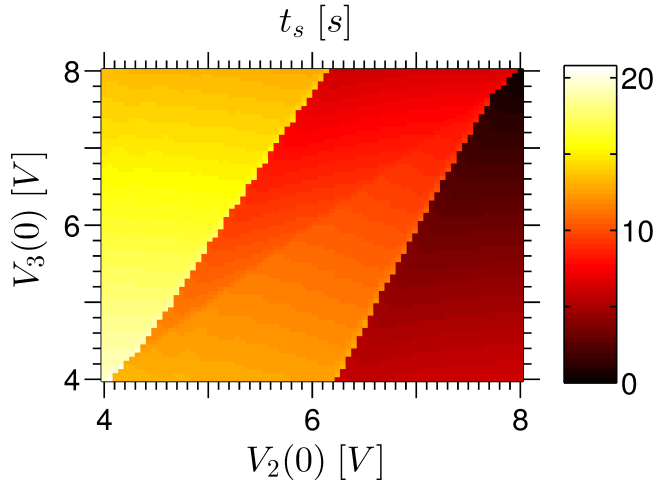}
 \end{center}
  \caption{Synchronization time as a function of the initial conditions of the LCOs $2$ and
  $3$, for fixed coupling strengths. The parameters of  the three LCOs are set identical
  ($\lambda = 26\;s^-1$ and $\gamma = 1050\;s^-1$), with coupling strengths $\beta_{1} = 4
  \;V/s$ and $\beta_{2} = 7\;V/s$  and initial condition of LCO 1 is fixed at $V_{01} =
  2V_{cc}/3$. Initial conditions of LCOs 2 and 3 are ranged from $V_{cc}/3$ to $2V_{cc}/3$
  with a step of $0.053\;V$. }
\end{figure}

In the case of configuration B [Fig. \ref{Configurations}(b)], LCO$_1$ and LCO$_3$ are in MI
configuration. The intermediary destroys the symmetry that occurs with $\beta_2=0$. It is
necessary again, that LCO$_1$ and LCO$_2$ synchronize before for the synchronization between
LCO$_1$ and LCO$_3$ is possible. Similar arguments to the analysis of the previous
configurations lead to the existence of some discontinuities between synchronization times
and coupling strengths for the configuration B as well.

Another case of interest is the situation where LCO$_2$ is not identical to LCO$_1$ and LCO$_3$. Such a system is capable of showing other types of synchronization.
For autonomous systems (such as three coupled LCO), generalized synchronization has been shown to be possible  \cite{alvarez2008generalized},
though synchronization, in general, is not guaranteed.  For the MS configuration with intermediate hierarchy, synchronization between
LCO$_1$ and LCO$_2$ is only possible if the frequency detuning and coupling intensity are such that LCO$_2$ falls within an Arnold Tongue of the LCO$_1$-LCO$_2$ interaction 
\cite{rubido2011synchronization}. If it falls within a 1-1 Arnold Tongue, total synchronization is guaranteed.
Also, for higher-order Arnold Tongues, LCO$_2$ will present generalized synchronization with respect to LCO$_1$, but then the synchronization of LCO$_3$  
with respect to LCO$_1$ is no longer guaranteed. The case of the MI configuration with non- identical intermediate hierarchy is far more complex and is left for further studies.            

Finally, we remark that we are able to derive conditions under which the inclusion of
intermediaries reduces the synchronization times between two LCOs. In the case of MS with
intermediary, the necessary condition to reduce the synchronization time with respect to the
two-LCO configuration is
\begin{equation}
 t_{s 1-2}+\mathrm{max}\left\{ t_{s 1,2-3}  \right\}_{\Delta \phi} <t_{s 1-3\left(free\right)},
\label{eq:condition}
\end{equation}
where $t_{s 1,2-3}$ is the synchronization time between LCO$_{1,2}$ and LCO$_3$, and
$t_{s 1-3\left(free\right) }$ is the synchronization time between LCO$_1$ and LCO$_ 3$ when
LCO$_ 2$ is absent. The second term on the left in  Eq.~\ref{eq:condition} is the maximum
value of $t_{s 12,3}$ among every possible pair of initial conditions $\phi_{1,2}^{i}$ and
$\phi_{3}^{i}$.

\section{Conclusions}

Groups of two and three coupled LCOs present a number of interesting synchronization
properties. This work includes results for identical oscillators, though further studies
indicate the validity of our results and analysis for groups of quasi-identical oscillators
and, in smaller regions, for larger frequency mismatches between the LCOs. Specifically,
we studied the influence of coupling strengths over synchronization times for groups of two
LCOs in MS and MI configurations, and for groups of three LCOs in a MS configuration with an
intermediate hierarchy and a MI configuration with an intermediate hierarchy. When comparing
synchronization times for two-LCO and three-LCO dynamics, we found that unlike systems of two
LCOs, synchronization times for systems of three LCOs do not always decrease when coupling
strengths increases, and larger non-synchronous regions are found. In three LCO dynamics, we
found an interesting, unexpected effect: a discontinuous relationship between synchronization
times and coupling strengths or initial conditions. Furthermore, we presented a qualitative
analysis of the dynamics of three LCOs coupled in configurations (a) and (b) of
Fig.~\ref{Configurations}, that justifies the existence of such discontinuities. In the light
of this analysis, we found sufficient conditions for intermediate hierarchy to reduce
synchronization times with respect to the analogue configurations without the intermediate
hierarchy. These findings could be relevant for the construction and interpretation of functional networks, such as those constructed from brain or climate data 
\cite{barreiro2011inferring,rubido2014exact}.

\bibliographystyle{epj}

\bibliography{/home/marti/Dropbox/bibtex/mybib}

\end{document}